\documentclass[8.5pt,twoside,twocolumn]{article}
\usepackage[super,sort&compress,comma]{natbib} 
\usepackage{mhchem}
\usepackage{times,mathptmx}
\usepackage{sectsty}
\usepackage{balance} 

\usepackage{graphicx} 
\usepackage{lastpage}
\usepackage[format=plain,justification=raggedright,singlelinecheck=false,font=small,labelfont=bf,labelsep=space]{caption} 
\usepackage{fancyhdr,float}
\usepackage{textcomp,color,setspace}
\usepackage{lineno}
\begin{document}

\thispagestyle{plain}
\fancypagestyle{plain}{
\fancyhead[L]{\includegraphics[height=8pt]{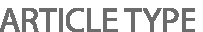}}
\fancyhead[C]{\hspace{-1cm}\includegraphics[height=20pt]{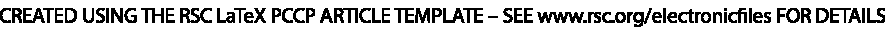}}
\fancyhead[R]{\includegraphics[height=10pt]{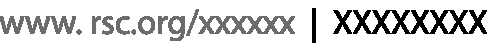}\vspace{-0.2cm}}
\renewcommand{\headrulewidth}{1pt}}
\renewcommand{\thefootnote}{\fnsymbol{footnote}}
\renewcommand\footnoterule{\vspace*{1pt}%
\hrule width 3.4in height 0.4pt \vspace*{5pt}} 
\setcounter{secnumdepth}{5}

\makeatletter 
\def\subsubsection{\@startsection{subsubsection}{3}{10pt}{-1.25ex plus -1ex minus -.1ex}{0ex plus 0ex}{\normalsize\bf}} 
\def\paragraph{\@startsection{paragraph}{4}{10pt}{-1.25ex plus -1ex minus -.1ex}{0ex plus 0ex}{\normalsize\textit}} 
\renewcommand\@biblabel[1]{#1}            
\renewcommand\@makefntext[1]%
{\noindent\makebox[0pt][r]{\@thefnmark\,}#1}
\makeatother 
\renewcommand{\figurename}{\small{Fig.}~}
\sectionfont{\large}
\subsectionfont{\normalsize} 

\fancyfoot{}
\fancyfoot[LO,RE]{\vspace{-7pt}\includegraphics[height=9pt]{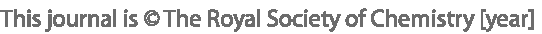}}
\fancyfoot[CO]{\vspace{-7.2pt}\hspace{12.2cm}\includegraphics{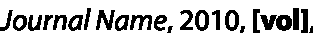}}
\fancyfoot[CE]{\vspace{-7.5pt}\hspace{-13.5cm}\includegraphics{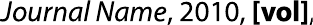}}
\fancyfoot[RO]{\footnotesize{\sffamily{1--\pageref{LastPage} ~\textbar  \hspace{2pt}\thepage}}}
\fancyfoot[LE]{\footnotesize{\sffamily{\thepage~\textbar\hspace{3.45cm} 1--\pageref{LastPage}}}}
\fancyhead{}
\renewcommand{\headrulewidth}{1pt} 
\renewcommand{\footrulewidth}{1pt}
\setlength{\arrayrulewidth}{1pt}
\setlength{\columnsep}{6.5mm}
\setlength\bibsep{1pt}

\twocolumn[
  \begin{@twocolumnfalse}
\noindent\LARGE{\textbf{Molecular Polymorphism: Microwave Spectra, Equilibrium Structures, and an Astronomical Investigation of the HNCS Isomeric Family}}
\vspace{0.6cm}

\noindent\large{\textbf{Brett A. McGuire,$^{a,b,\ddag}$ Marie-Aline Martin-Drumel,$^b$ Sven Thorwirth,$^c$ Sandra Br\"{u}nken,$^c$ Valerio Lattanzi,$^d$ Justin L. Neill,$^e$ Silvia Spezzano,$^d$ Zhenhong Yu,$^f$ Daniel P. Zaleski,$^e$ Anthony J. Remijan,$^a$ Brooks H. Pate,$^e$ and Michael C. McCarthy$^{\ast,a,g}$}}\vspace{0.5cm}

\noindent\textit{\small{\textbf{Received Xth XXXXXXXXXX 20XX, Accepted Xth XXXXXXXXX 20XX\newline
First published on the web Xth XXXXXXXXXX 200X}}}

\noindent \textbf{\small{DOI: 10.1039/b000000x}}
\vspace{0.6cm}
\normalsize{}

\doublespacing

The rotational spectra of thioisocyanic acid (HNCS), and its three energetic isomers (HSCN, HCNS, and HSNC) have been observed at high spectral resolution by a combination of chirped-pulse and Fabry-P\'{e}rot Fourier-transform microwave spectroscopy between 6 and 40~GHz in a pulsed-jet discharge expansion. Two isomers, thiofulminic acid (HCNS) and isothiofulminic acid (HSNC), calculated here to be 35-37~kcal/mol less stable than the ground state isomer HNCS, have been detected for the first time. Precise rotational, centrifugal distortion, and nitrogen hyperfine coupling constants have been determined for the normal and rare isotopic species of both molecules; all are in good agreement with theoretical predictions obtained at the coupled cluster level of theory. On the basis of isotopic spectroscopy, precise molecular structures have been derived for all four isomers by correcting experimental rotational constants  for  the effects of rotation-vibration calculated theoretically. Formation and isomerization pathways have also been investigated; the high abundance of HSCN relative to ground state HNCS, and the detection of strong lines of SH using CH$_3$CN and H$_2$S, suggest that HSCN is preferentially produced by the radical-radical reaction HS + CN.  A radio astronomical search for HSCN and its isomers has been undertaken toward the high-mass star-forming region Sgr B2(N) in the Galactic Center with the 100 m Green Bank Telescope.  While we find clear evidence for HSCN, only a tentative detection of HNCS is proposed, and there is no indication of HCNS or HSNC at the same rms noise level.  HSCN, and tentatively HNCS, displays clear deviations from a single-excitation temperature model, suggesting weak masing may be occurring in some transitions in this source.

\noindent \normalsize{}
\vspace{0.5cm}
 \end{@twocolumnfalse}
  ]


\doublespacing

\section{Introduction}

Isomerism is one of the oldest and most important concepts in chemistry, dating back to the 1820s when Liebig and W\"ohler first demonstrated that silver fulminate and silver cyanate  -- two compounds with the same elemental formula -- have different physical properties. These findings led Berzelius to propose  the  concept  of ``isomer'' in 1831.~\cite{esteban:1201}   Because it is fundamentally linked to molecular structure and chemical bonding, isomerism --- particularly of small molecules --- has long fascinated experimentalists and theoreticians alike.

Astrochemistry is one of the applied disciplines where structural isomers are of great importance because  chemistry in the interstellar medium (ISM) is kinetically, rather than thermodynamically, controlled.\cite{hirota:717}  Consequently, the abundances of isomers (e.g., HCN vs.~HNC)\cite{Loison:2014br}  in astronomical sources  often provide a sensitive probe of the chemical evolution and physical conditions that are operative there. In many cases, isomeric abundance ratios deviate significantly from predictions based on thermodynamic considerations alone; in some astronomical sources, a higher-energy isomer may even be more abundant than the most stable conformer.\cite{Lovas:2010ew,Zaleski:2013bc}  A recent illustration of this point is provided by studies of three stable singlet H$_2$C$_3$O isomers in the Sagittarius B2 (N), hereafter Sgr B2(N), star-forming region: Loomis et al.\cite{Loomis:2015jh} found no evidence for $l$-propadienone (H$_2$CCCO) in this source, but rotational lines of the higher energy isomer, cyclopropenone (calculated to lie 6~kcal/mol above ground),\cite{Zhou:2008un} are readily observed.  Their results show that $l$-propadienone is at least an order of magnitude less abundant than cyclopropenone.  Propynal [HC(O)CCH], a low-lying isomer of comparable stability to $l$-propadienone,\cite{komornicki:1652,maclagan:185,ekern:16109,Zhou:2008un,karton:22} is more than an order of magnitude more abundant than cyclopropenone.  The authors suggest that these large variations in abundance could arise from different formation pathways on the surface of interstellar dust grains, for which some supporting laboratory evidence has been found from surface reaction experiments.\cite{Zhou:2008un} 

\footnotetext{$^a$~National Radio Astronomy Observatory, 520 Edgemont Rd, Charlottesville, VA USA 22903.}
\footnotetext{$^b$~Harvard-Smithsonian Center for Astrophysics, 60 Garden Street, Cambridge, MA USA 02138.}
\footnotetext{$^c$~I. Physikalisches Institut, Universit\"{a}t K\"{o}ln, K\"{o}ln Germany 50937.}
\footnotetext{$^d$~Max-Planck Institut fur Extraterrestrische Physik, Garching, Bayern Germany.}
\footnotetext{$^e$~Department of Chemistry, University of Virginia, 759 Madison Ave, Charlottesville, VA 22903.}
\footnotetext{$^f$~Aerodyne Research, Inc, Billerica, MA USA.}
\footnotetext{$^g$~School of Engineering and Applied Sciences, Harvard University, 29 Oxford Street, Cambridge, MA USA 02138.}
\footnotetext{$^{\ddag}$~B.A.M. is a Jansky Fellow of the National Radio Astronomy Observatory.}

Because nearly all well-known interstellar species have high-energy isomers with significant barriers to either isomerization or dissociation, there is much applied interest in identifying and precisely characterizing the rotational spectra of these metastable forms. Isocyanic acid (HNCO) was one of the first polyatomic molecules detected in space,\cite{snyder:619,buhl:625} but owing to the lack of laboratory rest frequencies, it was not until fairly recently that its higher energy isomers cyanic acid, HOCN (24.7~kcal/mol above ground),\cite{Schuurman:2004je} and fulminic acid, HCNO (70.7~kcal/mol), were observed in interstellar molecular clouds.\cite{brunken:880,marcelino:l27} Soon after its initial astronomical identification,\cite{brunken:880} rotational lines of HOCN were reported in at least five other sources,\cite{brunken:2010hp,marcelino:a105} findings which suggest that this isomer and HCNO are common constituents of the ISM. Abundance ratios and formation pathways of the HNCO isomers as a function of visual extinction, density, and temperature have now been the subject of several chemical modeling studies.\cite{marcelino:a105,quan:2101,jimenez:19} Among the four singlet isomers, only isofulminic acid (HONC) has yet to be detected in space.

Apart from their astronomical interest, systematic studies of the structure, properties, and formation pathways of isomers are of fundamental importance. Such studies provide insight into a wide variety of bonding preferences (e.g., bond order), enable comparative studies of isovalent systems, and provide stringent tests for quantum chemical calculations.   For example, rotational spectroscopy measurements of the elusive HONC (84~kcal/mol) -- the highest energy singlet isomer of HNCO\cite{mladenovic:174308} -- are consistent with a structure containing a polar C-N triple bond\cite{poppinger:7806} and a significant HOC bending angle (105$^{\circ}$), in good agreement with a high-level coupled cluster calculation performed in conjunction with the experimental work.  Furthermore, simultaneous detection of several isomers under similar experimental conditions may yield insight into isomerization pathways, and provide estimates of relative and absolute abundances, so follow-up experiments can be undertaken at other wavelengths.

Like their isovalent oxygen counterparts, thiocyanates (R-SCN) and isothiocyanates (R-NCS) have a rich history which is closely linked to isomerization. In one of the first investigations of these compounds, Hofmann established in 1868 that methyl thiocyanate rearranges to form methyl isothiocyanate at high temperature.\cite{hofmann:201} Several years later, Billeter\cite{billeter:462} and Gerlich\cite{gerlich:80} independently observed that allyl thiocyanate (H$_2$C=CH-CH$_2$-NCS) thermally rearranges into allyl isothiocyanate, known as mustard oil,\cite{smith:3076} the compound largely responsible for the characteristic hot, pungent flavor of vegetables such as horseradish and mustard leaves. Today, the chemistry of thiocyanates and their derivatives is extensive and highly varied; compounds containing this chromophore are used in applications ranging from pharmaceuticals, dyes, and synthetic fiber to fungicides and photography, among others.

\begin{figure*}
\centering
\includegraphics[width=\textwidth]{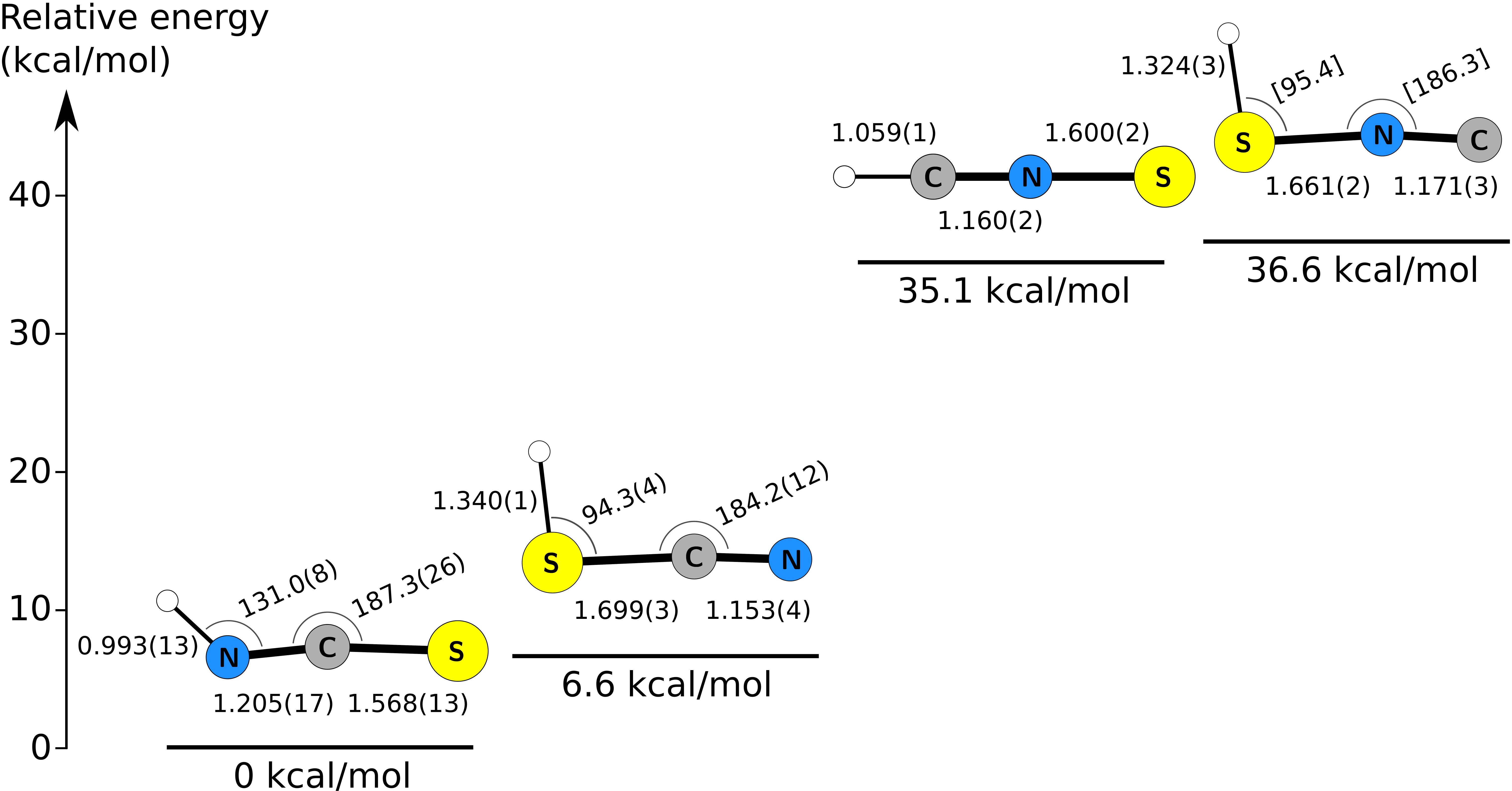}
\caption{Relative energies and structures of the [H, N, C, S] isomer family.  Energies are calculated at the 
ae-CCSD(T)/cc-pwCVQZ level and corrected for zero-point vibrational contributions calculated at the fc-CCSD(T)/cc-pV(Q+$d$)Z level. Semi-experimental equilibrium ($r_e^{SE}$) structures, obtained in this work, are indicated, along with associated uncertainties (1$\sigma$) derived from a least-squares optimization.  Bond lengths are in \r{A}, bond angles are in degrees. Square brackets indicate the structural parameter was fixed to the calculated value.}
\label{structures}
\end{figure*}

Isothiocyanic acid (HNCS), the simplest isothiocyanate, is calculated to be the most stable molecule in the [H, N, C, S] family, followed by thiocyanic acid (HSCN), lying about 6~kcal/mol above HNCS.\cite{bak:666,Wierzejewska:2003dd} These are followed by thiofulminic acid (HCNS) and isothiofulminic acid (HSNC), which are comparably stable to each other at 35.1~kcal/mol and 36.6~kcal/mol above HNCS, respectively (this work).  Each isomer is calculated to possess a singlet ground state with a nearly linear heavy atom backbone, a planar equilibrium structure, and a large permanent electric dipole moment.  Fig.~\ref{structures} summarizes a number of the salient properties of the four singlet HNCS isomers.

With the exception of HNCS itself, our knowledge of the [H, N, C, S] isomeric system is limited.  HNCS was first detected nearly 50 years ago by Beard and Dailey~\cite{Beard:1950ez} who determined its molecular structure and dipole moment from an analysis of the rotational spectrum and those of its more abundant isotopologues.  Since then, HNCS and its isotopic species have been the subject of many high-resolution studies, from the microwave to the far infrared region;\cite{yamada:189,Yamada:1980ky} much of this work was aimed at understanding how large-amplitude bending vibrations affect structural rigidity.  HNCS has also long been known as a constituent of the ISM; it was detected nearly 40 years ago via several $a$-type, $K_a=0$ rotational transitions toward the Sgr B2 region.\cite{Frerking:1979yd}

Until quite recently, there was little spectroscopic information on HSCN, HCNS, or HSNC.  HSCN and HSNC were first characterized experimentally at low spectral resolution by matrix-IR spectroscopy\cite{Wierzejewska:227} in which both isomers  were formed by UV-photolysis of HNCS in solid argon and nitrogen matrices, but HCNS does not appear to have been studied at any wavelength. The microwave spectrum of HSCN was recently reported by several co-authors of this study,\cite{Brunken:2009fh} and soon after it was detected in the ISM.\cite{Halfen:2009it,adande:561}  Owing to the absence of rotational spectra for HCNS and HSNC, astronomical searches had not been possible.

Since sulfur is less electronegative and has a larger atomic radius compared to oxygen, the energetics, structure, and bonding of analogous [H, N, C, S] and [H, N, C, O] isomers are predicted to differ.   While the [H, N, C, S] isomers have the same energy ordering compared to their oxygen counterparts, the spread in energy is more than two times smaller (37 vs.~84~kcal/mol).\cite{Schuurman:2004je,Wierzejewska:2003dd}  Because the reservoir of sulfur in space is not well established in dense molecular clouds,\cite{wakelam:159} the abundance of higher energy [H, N, C, S] isomers may significantly differ from that found in the HNCO isomers. 

In this paper we report a comprehensive laboratory study of the microwave spectra of HCNS and HSNC, the two remaining singlet isomers of HNCS, along with detection of their singly-substituted isotopic species and a number of rare isotopologues of HSCN, using both chirped-pulse Fourier-transform microwave (CP-FTMW) spectroscopy and cavity-FTMW spectroscopy. Because all four isomers can be simultaneously observed under the same experimental conditions, it has been possible to derive abundances relative to ground state HNCS, and consequently infer the dominant chemical reaction that yields HSCN. The relatively low abundance found for HSNC, with respect to isoenergetic HCNS, may indicate a low  barrier to isomerization.  By correcting  the experimental rotational constants of each isotopic species for the effects of zero-point vibration calculated theoretically, precise semi-experimental equilibrium structures ($r_e^{\rm SE}$) have been derived for each isomer.  Finally, the results of an astronomical search using observations toward Sgr B2(N) from the Green Bank Telescope (GBT) Prebiotic Interstellar Molecular Survey (PRIMOS) project are reported. Although the recent millimeter study by Halfen et al.\cite{Halfen:2009it} toward this source found that HNCS and HSCN are present in nearly equal column density, in the PRIMOS survey, firm evidence is found for HSCN alone, with only a tentative detection of HNCS. Lines of HSCN are observed in both absorption and emission, an indication that the excitation of this isomer is not well described by a single excitation temperature.


\section{Quantum Chemical Calculations}
\label{qcc}

Quantum chemical calculations guided the initial spectroscopic searches for HCNS and HSNC.  These were performed using the CFOUR suite of programs,\cite{cfour,harding_JChemTheoryComput_4_64_2008} and follow the general strategies outlined in Ref.~\cite{puzzarini_IntRevPhysChem_29_273_2010} Briefly, calculations were performed at the coupled-cluster singles, doubles, and perturbative triple excitations [CCSD(T)] level of theory,\cite{raghavachari_chemphyslett_157_479_1989} and Dunning's hierarchies of correlation-consistent polarized valence and polarized core valence basis sets.  In the frozen core (fc) approach, the tight-$d$-augmented basis sets cc-pV($X+d$)Z ($X$ = T and Q) were used for the sulfur atom, and the corresponding cc-pV$X$Z basis sets for nitrogen, carbon, and hydrogen.\cite{dunning_JCP_90_1007_1989,dunning_JCP_114_9244_2001} The cc-pwCV$X$Z ($X=$ T and Q) basis sets were used when considering all electrons in the correlation treatment.\cite{peterson_JCP_117_10548_2002}

Equilibrium geometries were calculated using analytic gradient techniques\cite{Watts:1992ff}  and basis sets as large as cc-pwCVQZ,\cite{Peterson:2002cf} a level that has been shown to yield highly accurate molecular equilibrium structures even for molecules containing second-row elements.\cite{Coriani:2005cz,thorwirth_JCP_2009,mueck_AlCCH_2015,martin-drumel_JCP_144_084305_2016} Dipole moment components and nuclear quadrupole coupling constants were derived at the same level. Vibrational effects were calculated at the fc-CCSD(T)/cc-pV(Q+\emph{d})Z level using second-order vibrational perturbation theory (VPT2) based on the formulas given in Ref.\cite{mills_alphas} Harmonic force-fields were computed analytically, \cite{gauss_chemphyslett_276_70_1997} while cubic and semi-diagonal quartic force fields were obtained via numerical differentiation of the analytically evaluated harmonic force fields.\cite{stanton_JCP_108_7190_1998} Overall, these calculations provided harmonic vibrational frequencies, centrifugal-distortion and vibration-rotation interaction constants ($\alpha_i$), zero-point vibrational corrections to rotational constants ($\Delta B_0=\sum_i \alpha_i^B(d_i/2)$) as well as fundamental vibrational frequencies ($\nu_i$). Best estimates for the ground state rotational constants $A_0$, $B_0$, and $C_0$ were then obtained using the relation $B_0=B_e-\Delta B_0$ (with similar equations for $A_0$ and $C_0$) where the equilibrium rotational constants $A_e$, $B_e$, and $C_e$ are calculated from
the ae-CCSD(T)/cc-pwCVQZ structure.


\section{Experimental Methods}

Although metastable HCNS and HSNC are calculated to lie $\sim$35~kcal/mol above the HNCS global minimum (Fig.~\ref{structures}), prior work has shown that high-lying metastable isomers can be readily produced with an electrical discharge source and then probed in a rotationally-cold supersonic expansion.\cite{McCarthy:2016eo}  This unusual scenario arises because the reaction chemistry is closely coupled to the free energy of the electrons, which have an average kinetic energy of several~eV,\cite{Sanz:2005kc} rather than to the extremely low ($\sim$3 K) rotational excitation temperature of the expansion.  The isovalent [H, C, N, O] isomers to those sought here, HCNO and HONC, lie substantially higher in energy relative to the ground state HNCO, (70.7 and 84.1~kcal/mol, respectively\cite{Mladenovic:2009im}) yet they were readily observed using essentially the same production and detection method as the one described here.\cite{Schuurman:2004je,Brunken:2009ew,Mladenovic:2009hd,Mladenovic:2009im}


\subsection{Chirped-Pulse FT Microwave Spectroscopy}

Two configurations of the CP-FTMW spectrometer, one operating between 7.5 and 18.5 GHz ($K$-band) and the other between 25 and 40 GHz ($K_a$-band), were employed in this study.  Their design and operation have been previously reported\cite{Brown:2008gk,Zaleski:2012df}.  Briefly, short ($\sim1$~{\textmu}s) frequency chirps are amplified by either a 300~W (7.5--18.5 GHz) or a 50~W (25--40 GHz) traveling wave tube amplifier and broadcast into a vacuum chamber by a standard gain horn antenna.  Two or three pulsed nozzles, for the higher- and lower-frequency spectrometers, respectively, each equipped with an electrical discharge source,\cite{McCarthy:2000ti} were positioned perpendicular to the axis of the microwave propagation in such a way that the resulting supersonic expansion of each intersects with the microwave radiation about 10~cm from the valve orifice.  Because the transmission efficiency between the two horn antennas is poorer at $K_a$ band compared to $K$ band, the horns must be placed closer to each other in the high-band spectrometer, an arrangement which precludes using a third nozzle.

Ten frequency chirps,  separated by 30~{\textmu}s, are applied during each valve pulse. The resulting molecular free induction decay (FID) associated with each chirp is collected for 20~{\textmu}s  by a second matched horn antenna, and the signal is amplified using a standard low-noise amplifier.  The FID is then digitized at 100~GSa/s on a high speed (20~GHz) digital oscilloscope; the Fourier-transform of the FID yields the frequency-domain spectrum.



\begin{figure*}
\centering
\includegraphics[width=\textwidth]{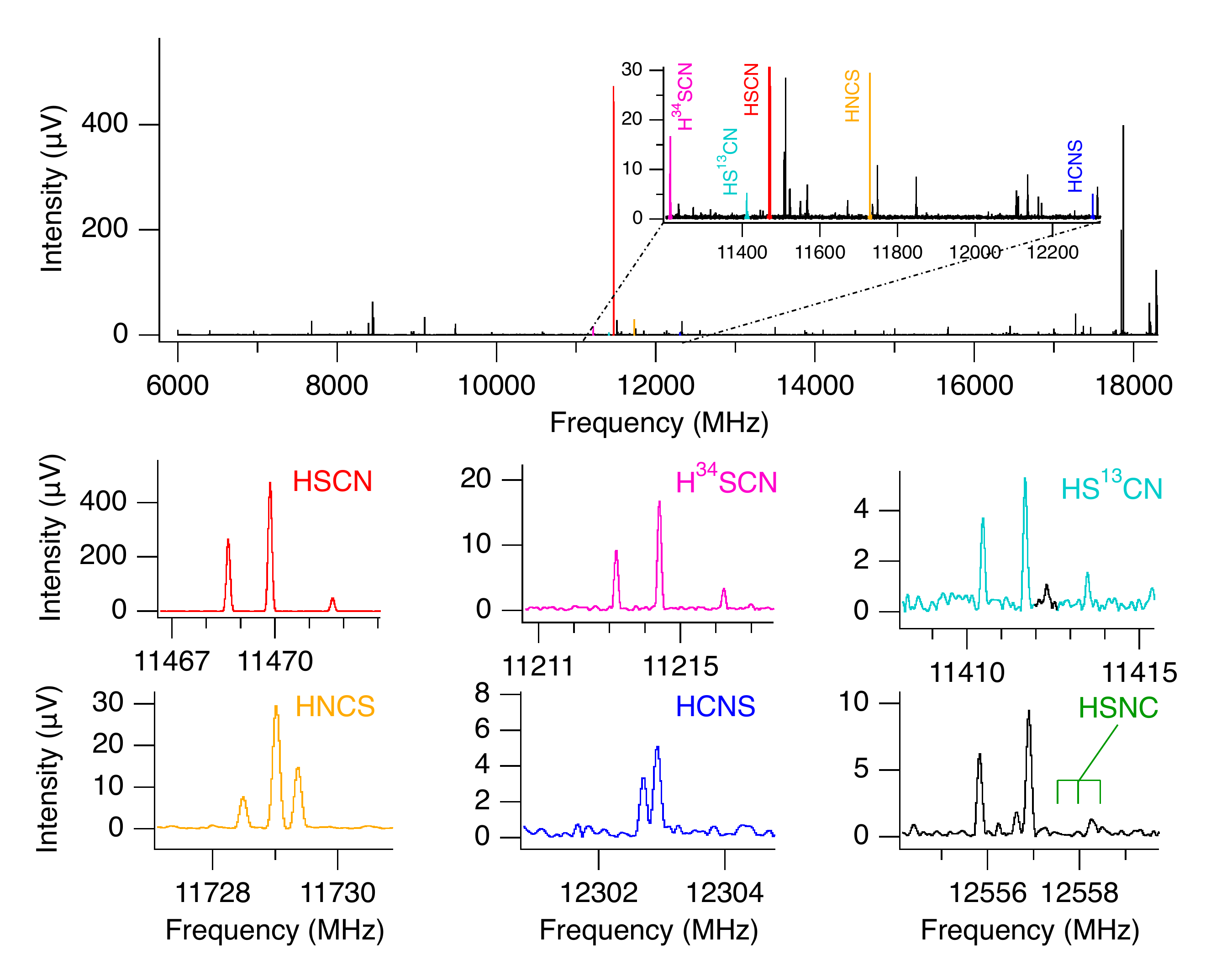}
\caption{Chirped pulse spectrum between 7.5 and 18.5 GHz through a dilute discharge of H$_2$S and CH$_3$CN.  The full bandwidth is shown, along with zoomed insets. The strongest feature in the spectrum arises from HSCN, and, owing to the high SNR, the $^{34}$S and $^{13}$C isotopologues are observed in natural abundance.  The fundamental rotational transitions of each species are shown in colors.  The hyperfine-split line of HSNC, not detected in this spectrum, is also indicated.  The strong transitions above $\sim$17.9 GHz are from CH$_3$CN and its isotopologues.}
\label{chirpspectra}
\end{figure*}

The discharge conditions were optimized to produce HSCN in high yield: a mixture of hydrogen sulfide (H$_2$S, 0.15\%) and methyl cyanide (CH$_3$CN, 0.15\%) heavily diluted in neon, a 1500 V dc discharge, and a stagnation pressure of 2.5~kTorr (3.2~atm) behind the 1 mm diameter supersonic nozzle were found to be optimal.  Using these conditions, chirped spectra were collected over a 12-hour period in each frequency band.  At the valve repetition rate of 5~Hz, nearly 2 million FIDs were recorded. The resulting spectrum has a high signal-to-noise ratio (SNR), and consequently the $^{34}$S and $^{13}$C isotopologues of HSCN were observed in natural abundance (see Figure \ref{chirpspectra}) in addition to the parent isotopologues of HNCS, HSCN, and HCNS.    At this level of integration, about 500 molecular lines were detected in the two bands.  In the high-band spectrum, for example, 24 molecules were readily identified: 18 are known interstellar molecules, a number which represents roughly 10\% of known astronomical species (Table S1).  In addition to the work on the [H, N, C, S] family presented here, an analysis of this spectrum led to the first interstellar detection of $E$- and $Z$-ethanimine (CH$_3$CHNH).\cite{Loomis:2013fs}

Because HSCN and its isomers are relatively light molecules, a single rotational transition falls in each of the two CP-FTMW bands, although a third transition of HSNC falls near the lower edge of the $K_a$ band.   The CCSD(T) calculations performed here, however, are generally accurate to a few 0.1\%, and zero-point energy corrections to the $B_{e}$ constants are generally small ($\sim0.4\%$) as well, meaning that the frequency uncertainty for the fundamental rotational transition of each isomer ($1_{0,1} \rightarrow 0_{0,0}$, predicted near 12~GHz) is expected to be less than $\pm$75 MHz.  Combined with the low line density and characteristic hyperfine splitting from the $^{14}$N nucleus, there were very few candidates lines for HCNS in the CP-FTMW spectrum.  The lack of any viable candidate lines for HSNC, however, necessitated a follow-up study using more sensitive cavity FTMW spectroscopy.

\subsection{Cavity FT Microwave Spectroscopy}

Following the CP-FTMW measurements, a number of follow-up studies were performed using a cavity FTMW spectrometer operating between 5 and 43~GHz. These studies were needed to: (1) confirm the initial identifications with better SNR and at higher spectral resolution; (2) extend the rotational spectroscopy to higher-$J$ transitions for the purpose of determining the leading centrifugal distortion terms; (3) detect rare isotopic species of isomers, both to provide overwhelming evidence for the identification and to allow precise determinations of molecular structures; and (4) detect HSNC, whose lines were not readily apparent in the CP-FTMW spectra.

Details of the cavity FTMW spectrometer, which is  based on the original design of Balle \& Flygare,\cite{Balle:1981ex} and its operation have been published previously.\cite{McCarthy:2000ti,Crabtree:2013fv,Grabow:2005kf}  The nozzle source is identical to that used in the CP-FTMW experiment, except this source is mounted  directly behind one of the cavity mirrors, and gas expands into the open resonator via a small hole in that mirror.  As the gas passes through the beam waist of the cavity, molecules are excited by a short (1~{\textmu}s) pulse of microwave radiation.  The resulting FID is recorded using a sensitive microwave receiver; the Fourier-transform of the FID yields the frequency-domain spectrum.  Owing to the high $Q$ of the cavity, the instantaneous bandwidth is only about 1~MHz, but the sensitivity per unit MHz and time is substantially higher compared to the CP-FTMW spectrometer, by roughly a factor of 40.\cite{Brown:2008gk}  The spectral resolution, determined by the divergence of the molecular beam,~\cite{Gottlieb:2000nc} is also quite high, approaching 0.1 ppm, or about a factor of 50 greater than that which can routinely be achieved in CP-FTMW spectra at the same frequency. 

Owing to slight differences in nozzle geometry and expansion conditions with the chirped-pulse experiment, the optimum production conditions for HNCS and HSCN were similar but not identical between the two experiments.  Although both molecules were observed using a mixture of H$_2$S  and CH$_3$CN, a dilute mixture of H$_2$S (0.035\%) and cyanogen (NCCN, 0.03\%) in neon and a discharge voltage of 800 V yielded somewhat stronger lines at the same backing pressure (2.5~kTorr) behind the nozzle. 

To confirm the initial CP-FTWM assignments, searches were undertaken at higher frequencies to establish if additional lines were present at frequencies corresponding to ratios of integers, i.e.~near 24 and 36~GHz for HSCN, as all four [H, N, C, S] isomers are either linear molecules or asymmetric tops very close to the prolate limit ($\kappa \sim 1$). Two series of lines, both observed with good SNR, were quickly identified in this manner.  Transitions of HSNC, undetected in the chirped-pulse experiment, were identified near 13, 25, and 38 GHz.  Many additional cavity measurements were performed to detect isotopic species of HCNS, HSNC, and HSCN, either in natural abundance ($^{34}$S) or using a variety of isotopically-enriched precursors (D$_2$S, CH$_3^{13}$CN, CH$_3$C$^{15}$N), in which searches were guided by the quantum chemical calculations described in Sec. 2.


\section{Experimental Results and Spectroscopic Analysis}

Spectroscopic constants of all species were determined by least-squares optimization using the CALPGM (SPFIT/SPCAT) suite of programs\cite{Pickett:1991cv} and either a standard asymmetric top ($S$-reduced) or linear molecule Hamiltonian, both of which include centrifugal distortion and hyperfine interactions, either nuclear quadrupole coupling from the $I=1$ spin of nitrogen, deuterium, or both, or spin-rotation from the $I=1/2$ $^{13}$C nucleus.  Frequencies derived from the higher resolution cavity FTMW data were used in each analysis.

\subsection{Isotopic HSCN (Thiocyanic Acid)}

HSCN is an asymmetric rotor very close to the prolate limit ($\kappa$ = $-$0.999)\cite{Townes:1975ve}, with an $a$-inertial axis almost coincident with the nearly-linear SCN heavy atom backbone, a large energy separation ($E/k \sim 15$ K) between successive $K_a$ levels, and a strong $a$-type spectrum.  Because it is produced in high abundance in the discharge nozzle, it was possible to detect transitions in the $K_a = 1$ ladders for several isotopic species, despite the very low rotational excitation temperature. The $K_a = 1$ lines, however, are a factor of 1000 weaker in intensity than those from the same $J$ but in the corresponding $K_a = 0$ ladder.

The rotational spectra of five isotopic species, H$^{34}$SCN, HS$^{13}$CN, HSC$^{15}$N, D$^{34}$SCN, and DSC$^{15}$N, were observed in the centimeter-wave band (Tables S2 and S3) and analyzed in this study.  Because the present data set is limited to $a$-type transitions with $K_a \leq 1$ below 40\,GHz, it was not possible to determine the $A$ rotational constant to better than $\sim$3\% for the deuterated isotopic species, and even less so for H$^{34}$SCN, HS$^{13}$CN, and HSC$^{15}$N. Consequently, the $A$ rotational constant of these species was fixed to the value derived for the main isotopologue in the fits.  Table \ref{hscnconstants} provides a summary of the derived constants for isotopic HSCN.

\begin{table*}
\centering
\caption{Experimental spectroscopic constants of ground state HSCN and its isotopic species (in MHz).}
\label{hscnconstants}
\footnotesize
\begin{tabular}{c cc c c c c c}
\hline
Constant$^a$					&	HSCN$^b$ 		&	H$^{34}$SCN	&	HS$^{13}$CN	&	 HSC$^{15}$N 		& 	DSCN$^b$		&	D$^{34}$SCN		&	DSC$^{15}$N	\\
\hline
$A_0$						&	289737(64) 		& 	[289737]$^c$	&	[289737]$^c$ 	&	[289737]$^c$		&	151350(478)		&	148302(568)		&	150423(1051)	\\ 
$B_0$						&	5794.71368(20)  	&  	5664.4752(4)	&	5765.0305(5)	&	5705.0993(3)		& 	5705.09910(29) 	&	5583.8130(3)		&	5495.9129(5)	\\
$C_0$						&	5674.93940(20) 	&	5549.7261(3)	&	5646.4473(3)	& 	[5470.1341]$^d$ 	&	5489.24957(27)	&	5376.0826(3)		&	5295.2994(5)	\\
$10^3D_J$					& 	1.66557(21)		&	1.61(2)		&	1.62(2)		& 	 [1.6648]$^c$		&	1.560(20)			&  	1.36(3)			&	1.37(4)		\\
$D_{JK}$						& 	0.15153(11) 		&	0.1440(2)		&	0.1484(2)		&	 [0.1515]$^c$		&	0.14152(34)  		&	0.1372(4)			&	0.1331(6)		\\
$10^6d_1$              				& 	$-$35.91(21)		&	$\cdots$ 		&	$\cdots$  		&	$\cdots$ 			&	 $\cdots$ 			& 	$\cdots$			&	$\cdots$		\\
$\chi_{aa}$(N)					&	$-$4.0477(15) 		& 	$-$4.047(2)	&	$-$4.049(2)	& 	$\cdots$ 			&	$-$4.0530(11)		&	$-$4.051(1)		&	$\cdots$		\\
$\chi_{bb}$(N)					&	2.8271(18) 		& 	2.818(3)		&	2.825(3)		& 	$\cdots$			&	2.8292(18)		&	2.829(4)			&	$\cdots$		\\
$\chi_{aa}$(D)					& 	$\cdots$ 			&	$\cdots$		&	$\cdots$		& 	$\cdots$			&	$-$0.0821(32)		&	$-$0.068(3)		&	$-$0.058(2)	\\
$\chi_{bb}$(D)					&	$\cdots$			&	$\cdots$ 		&	$\cdots$ 	 	& 	$\cdots$			&	0.1445(26) 		&	0.134(3)			&	0.136(3)		\\
\vspace{0.5em}	\\				
rms (kHz)						&	$\cdots$			&	2.2			&	2.8			&	4.8				&	$\cdots$			&	1.9				&	1.6			\\
$N_{lines}$					&	$\cdots$			&	27			&	24			&	3				&	$\cdots$			&	66				&	25			\\
\hline
\end{tabular}\\
\textsuperscript{\emph{a}} Values in parentheses are 1$\sigma$ uncertainties in units of the last significant digit.\\
\textsuperscript{\emph{b}} Ref.~\cite{Brunken:2009fh}.  Additional higher-order constants, provided in Ref.~\cite{Brunken:2009fh}, are not reproduced here.\\
\textsuperscript{\emph{c}} Fixed to value derived for normal HSCN.\\
\textsuperscript{\emph{d}} Owing to the limited data set, $B-C$ value constrained to value derived for normal HSCN.\\
\end{table*}

\subsection{HCNS (Thiofulminic Acid)}

The three lowest rotational transitions of HCNS and its four singly-substituted species H$^{13}$CNS, HC$^{15}$NS, HCN$^{34}$S, and DCNS have been measured between 12 and 36~GHz (see Tables~S4--S6).  Because this isomer is predicted to be strictly linear, for the normal isotopic species, HC$^{15}$NS, and HCN$^{34}$S, a fit rms comparable to the experimental uncertainty ($\sim$2~kHz) was achieved by optimizing only three parameters: the rotational constant $B_0$, the centrifugal distortion constant $D_0$, and the nitrogen quadrupole constant $eQq$(N) (except in the case of the $^{15}$N species). Owing to additional hyperfine structure from the deuteron and $^{13}$C for DCNS and H$^{13}$CNS, one additional hyperfine term, $eQq$(D) and the spin-rotation constant $C$($^{13}$C), respectively, was required to achieve a similarly low rms.  The best-fit constants are summarized in Table~\ref{hcnsconstants}, along with results from the quantum-chemical calculations from Sec.~2.  When comparison is possible, the agreement between the two sets of constants is excellent.

\begin{table*}
\centering
\caption{Experimental and calculated spectroscopic constants of ground state HCNS and its isotopic species (in MHz).}
\footnotesize
\label{hcnsconstants}
\begin{tabular}{ccccccc}
\hline
		&	\multicolumn{2}{c}{HCNS}			&	&	&  &	\\
\cline{2-3}
Constant$^a$		&	Measured			&	Calculated$^b$	&	H$^{13}$CNS	&	HC$^{15}$NS	&	HCN$^{34}$S	&	DCNS		\\
\hline
$B_0$			&	6151.4436(4)		&	6153.961		&	5931.3801(4)	&	6125.714(1)	&	6008.8818(6)	&	5617.8801(3)	\\
10$^3D_0$		&	1.71(2)			&	1.664		&	1.54(3)		&	1.71$^c$		&	1.71$^c$		&	1.38(2)		\\
$eQq$(N)			&	$-$0.723(2)		&	$-$0.711		&	$-$0.728(2)	&	$\cdots$		&	$-$0.723$^c$	&	$-$0.725(3)	\\
10$^3C$($^{13}$C)	&	$\cdots$			&	$\cdots$		&	11(1)			&	$\cdots$		&	$\cdots$		&	$\cdots$		\\
$eQq$(D)			&	$\cdots$			&	$\cdots$		&	$\cdots$		&	$\cdots$		&	$\cdots$		&	0.1958(4)		\\
\vspace{0.5em}	\\				
rms (kHz)			&	1.7				&	$\cdots$		&	2.4			&	$\cdots$		&	1.1			&	2.5			\\
$N_{lines}$		&	13				&	$\cdots$		&	19			&	3			&	3			&	29			\\
\hline
\end{tabular}\\
\textsuperscript{\emph{a}} Values in parentheses are 1$\sigma$ uncertainties in units of the last significant digit.\\
\textsuperscript{\emph{b}} See text for details.\\
\textsuperscript{\emph{c}} Fixed to value derived for normal HCNS.\\
\end{table*}

\subsection{HSNC (Isothiofulminic Acid)}

Similarly to HSCN, isothiofulminic acid is also an asymmetric rotor very close to the prolate limit ($\kappa= -0.999$) with the least principal inertial axis almost coincident with the nearly-linear SNC heavy atom backbone (Fig. \ref{structures}). Consequently, the molecule has a large $A$ rotational constant ($A=290$~GHz) and significant energy separation between successive $K_a$ energy levels.  At the very low rotational excitation temperatures that characterizes our supersonic jet, however, only rotational levels in the $K_a=0$ ladder are significantly populated.  As a consequence, only the effective constant $B_{eff} = B+C$ was determinable.  As with HCNS, the rotational spectra of the four singly-substituted isotopic species, H$^{34}$SNC, HS$^{15}$NC, HSN$^{13}$C, and DSNC,  have also been measured below 40 GHz (Tables S7 and S8). Best-fit constants are given in Table \ref{hsncconstants}.

\begin{table*}
\centering
\caption{Experimental and calculated spectroscopic constants of ground state HSNC and its isotopic species (in MHz).}
\label{hsncconstants}
\footnotesize
\begin{tabular}{c c  c c c c c}
\hline
		&	\multicolumn{2}{c}{HSNC}			&	&	&  &	\\
\cline{2-3}
Constant$^a$  			&	Measured		&	Calculated$^b$		&	H$^{34}$SNC		& 	HS$^{15}$NC	&	HSN$^{13}$C	&	DSNC			\\
\hline
$B_{eff}$				&	6279.0335(4)	&	6274.1645		&	6141.6485(6) 		&	6242.520(1)	&	6022.3790(3)	&	6118.1242(4)		\\
10$^3D_{eff}$			&	4.51(2)		&					&	[4.51]$^c$			&	4.4(2)	 	& 	3.95(2)		&	16.14(3)			\\
$\chi_{aa}$(N)			&	1.213(2)		&	1.204			&	[1.21]$^c$			&				&	1.214(2)		&	[1.21]$^c$			\\
$\chi_{aa}$(D)			&				&					&					&				&				&	$-$0.05(1)			\\
\vspace{0.5em}	\\				
rms (kHz)				&	1.6			&	$\cdots$			&	1.6				&	$\cdots$		&	2.5			&	2.2				\\
$N_{lines}$			&	13			&	$\cdots$			&	3				&	8			&	19			&	19				\\
\hline
\end{tabular}\\
\textsuperscript{\emph{a}} Values in parentheses are 1$\sigma$ uncertainties in units of the last significant digit.\\
\textsuperscript{\emph{b}} See text for details.\\
\textsuperscript{\emph{c}} Fixed to value derived for normal HSNC.\\
\end{table*}


\subsection{Structural Determinations}

Purely experimental ($r_0$) structures for HCNS, HSNC, and HSCN were derived by least-squares optimization of the structural parameters of each isomer to reproduce the measured moments of inertia of its normal and rare isotopic species  using the STRFIT program.\cite{Kisiel:2001es} Because the moment of inertia is inversely proportional to the rotational constant along each axis, the three moments of inertia of each species are trivially calculated from constants in Tables \ref{hscnconstants}, \ref{hcnsconstants}, and \ref{hsncconstants}.   For HNCS, no attempt was made to derive an $r_0$ structure because a substitution ($r_s$) structure has been previously reported.\cite{Yamada:1980ky}  For HCNS, only the three bond lengths were varied because of its linear geometry, but for the asymmetric top HSCN, the two bond angles were optimized as well. Owing to the limited data set for HSNC, the three bond lengths were varied, with the remaining two angles constrained to the calculated value.  For isotopic species, poorly constrained constants in the fit were omitted in the structural determination.  Tables S10--S13 summarize the geometries derived by this method.

Semi-experimental equilibrium ($r_e^\text{SE}$) structures were determined by correcting the experimental rotational constants ($A_0$, $B_0$, $C_0$) to account for zero-point vibrational effects as obtained by VPT2 (see Sec. \ref{qcc}).  The semi-experimental equilibrium rotational constants ($A_e^\text{SE}$, $B_e^\text{SE}$, and $C_e^\text{SE}$) are used instead of the experimental constants in the structural optimization.  The remainder of the procedure is otherwise identical to that employed for the $r_0$ structures.  For completeness, a $r_e^\text{SE}$ structure has also been derived for ground state HNCS, in which the published constants are corrected for zero-point vibration.\cite{Beard:1950ez,Niedenhoff:1995tf}    Vibrational corrections for each isotopic species of each HNCS isomer are summarized in Tables S14--S17, along with the corresponding $B_e^\text{SE}$ constants.  The best-fit $r_e^\text{SE}$ structures are given in Tables~S10--S13, and the values are reported in Fig. \ref{structures}.

For most isomers, the most obvious difference between the $r_0$ and $r_e^\text{SE}$ structures is the higher accuracy of the derived parameters in the latter structure.  For HSCN, the fractional precision of most parameters is improved by a factor of three or more, with bond lengths derived to a few m\AA, while the bond angles are accurate to 1$^{\circ}$ or better. The inertial defects derived for each isotopologue decrease by roughly a factor of 10 or more (e.g. from 0.0958 to 0.0045 amu \AA$^2$ for HSCN).  Despite being the lowest energy and best studied of the isomers, the structural parameters for HNCS are the least well-determined of the four due to significant influence of large-amplitude motions on the structure.\cite{Mladenovic:2009ca}

Compared to the analogous [H, N, C, O] isotopologues, the [H, N, C, S] species display longer bond lengths, by $\sim$0.4--0.5\,\AA, for bonds involving sulfur rather than oxygen.  CN, CH, and NH bond lengths are remarkably similar, usually within 0.01\,\AA.  The bond angles for which sulfur is the central atom are smaller than their O-atom counterparts by 10--15$^{\circ}$.\cite{Mladenovic:2009ca}

\subsection{Relative Abundances and Formation Pathways}

Table~\ref{lababundances} summarizes the abundances of HSCN, HCNS, and HSNC relative to HNCS in the chirped-pulse spectrum, which have been derived from intensities of the fundamental $a$-type rotational transition of each species, accounting for differences in the hyperfine structure and dipole moment. This calculation also assumes that line intensity is proportional to the square of the $a$-type dipole moment, that all four isomers have the same rotational excitation temperature in the expansion, and that the instrument response function  is uniform over the range of measurement (11.4--12.5~GHz).

\begin{table}
\centering
\caption{Calculated relative energies (in kcal/mol) and dipole moments (in Debye) of the four [H, N, C, S] isomers, and measured relative abundances in the CP-FTMW spectrum.}
\label{lababundances}
\begin{tabular}{ccccc}
\hline
		&		&	\multicolumn{2}{c}{Dipole moment\textsuperscript{\emph{a}}}		 &	\\
\cline{3-4}
Isomer		&	$E_{rel}$ \textsuperscript{\emph{a}}	&	$\mu_a$	&	$\mu_b$&	$N_X/N_{HNCS}$\textsuperscript{\emph{b}}		\\
\hline
HNCS		&	0					    &	1.64					& 1.16  &	1					\\
HSCN		&	6.6  					&	3.29					&  0.89      &	4					\\
HCNS		&	35.1					&	3.85					&        &	0.07		\\
HSNC		&	36.6					&	3.13					&  0.88     &	$<$0.004	\\
\hline
\end{tabular}\\
\textsuperscript{\emph{a}} Calculated at the ae-CCSD(T)/cc-pwCVQZ level of  theory, see text.\\
\textsuperscript{\emph{b}} Measurements made by comparing observed intensities of the fundamental rotational  transition for each species, and correcting for differences in dipole moment.  Errors are estimated to be 10-15\%.\\

\end{table}

Under our discharge conditions, HSCN is the most abundant isomer by a factor of four, even though this metastable isomer lies 6.6~kcal/mol above HNCS.  Because $\Lambda$-doublet transitions of the SH radical are readily observed near 8.4~GHz\cite{Meerts:1975kq}, and the CN radical is known to be a common fragment of CH$_3$CN dissociation,\cite{McDowell:1952dt} HSCN is likely produced in high abundance via the radical-radical reaction HS ($^2\Pi$) + CN ($^2\Sigma^+$).\cite{Wierzejewska:2003dd}  Although CN has no transitions in the frequency range of the CP-FTMW spectrum, detection of rotational lines of several other cyanides (e.g., HC$_3$N, CH$_2$CHCN, CH$_3$CH$_2$CN) and isocyanides (e.g., CH$_3$NC, HCCNC, CH$_2$CHNC; see Table~S1) strongly suggest that this radical plays an important role in the discharge chemistry. 

The counterpart of HSCN, isothiofulminic acid HSNC is roughly 10$^3$ times less abundant in the same discharge, and underabundant by a factor of 40 compared to HCNS, its isoenergetic isomer. Previous theoretical work has explored the unimolecular isomerization of HNCS isomers at the MP2/aug-cc-pVTZ level of theory for structures, and CCSD(T)/aug-cc-pVTZ level of theory for single-point energy calculations.\cite{Wierzejewska:2003dd}  Although formation of both HSCN and HSNC from HS + CN is exothermic and barrierless, HSCN resides in a deeper potential well with larger barriers to isomerization compared to that of HSNC. The HSNC to HSCN isomerization proceeds through a cyclic intermediate, S(H)CN, with an energy barrier corrected for zero-point energy of only 23~kcal/mol, while the reverse reaction must overcome a barrier of nearly 50~kcal/mol; a similarly high barrier is also required to convert HSCN to HNCS (see Fig. 5 in \citet{Wierzejewska:2003dd}) .  For these reasons, energized HSCN is apparently easier to stabilize  compared to energized HSNC and, once formed, HSCN probably undergoes relatively little isomerization. Because the exothermicity of the HS + NC $\rightarrow$ HSNC reaction is roughly three times the well depth of the HSNC minimum, however, considerable rearrangement to the thermodynamically-favored HSCN may occur prior to stabilization of HSNC.

Guided by the theoretical work of \citet{Wierzejewska:2003dd}, it appears likely that HNCS and HCNS are formed through reactions involving sequential reactions of CN with atomic S and H, either H ($^2$S) + NCS ($^2\Pi$), which can produce both HNCS and HSCN, or H ($^2$S) + CNS ($^2\Pi$), which only yields HCNS. Once formed, like HSCN, considerable barriers must be overcome to isomerize HNCS, the lowest of which is about 60~kcal/mol to form HSCN.  The preferential production of HCNS relative to HSNC, despite their nearly identical energetics, is a likely indication of much greater stability of the former to isomerization.  HCNS isomerization to either HSCN or HNCS is energetically unfavorable; both involve barriers of about 60~kcal/mol.

\section{Astronomical Search}

We have conducted a search for each of the four [H, N, C, S] isomers in the publicly-available PRIMOS survey\footnote{The entire PRIMOS dataset, as well as further information on the project, is accessible at http://www.cv.nrao.edu/PRIMOS/.} toward Sgr B2(N)\cite{Neill:2012fr} where HSCN and HNCS have been previously detected.\cite{Halfen:2009it,Frerking:1979yd,Belloche:2013eba}  Based on a purely thermodynamically-driven treatment, a detectable column density of HSNC or HCNS in this source is not expected, given their much lower calculated stabilities.  However, prior observations of Sgr B2(N) have shown that chemistry in this star-forming region is kinetically- rather than thermodynamically-controlled,\footnote{The reaction rates and barriers, both to formation and destruction, of molecules almost always dominate over difference in their relative energies (stabilities).} and abundances of higher-energy isomers commonly defy thermodynamic predictions.\cite{Loomis:2015jh,Neill:2012fr}  Sgr B2(N) is a spatially- and radiatively-complex source, and careful treatment is required to obtain accurate column densities and excitation temperatures.  Here, we follow the convention of \citet{Hollis:2004uh} to derive these values; a detailed description of the techniques is provided in the Supplemental Information.

\subsection{HNCS}

\begin{figure}
\centering
\includegraphics[width=0.5\textwidth]{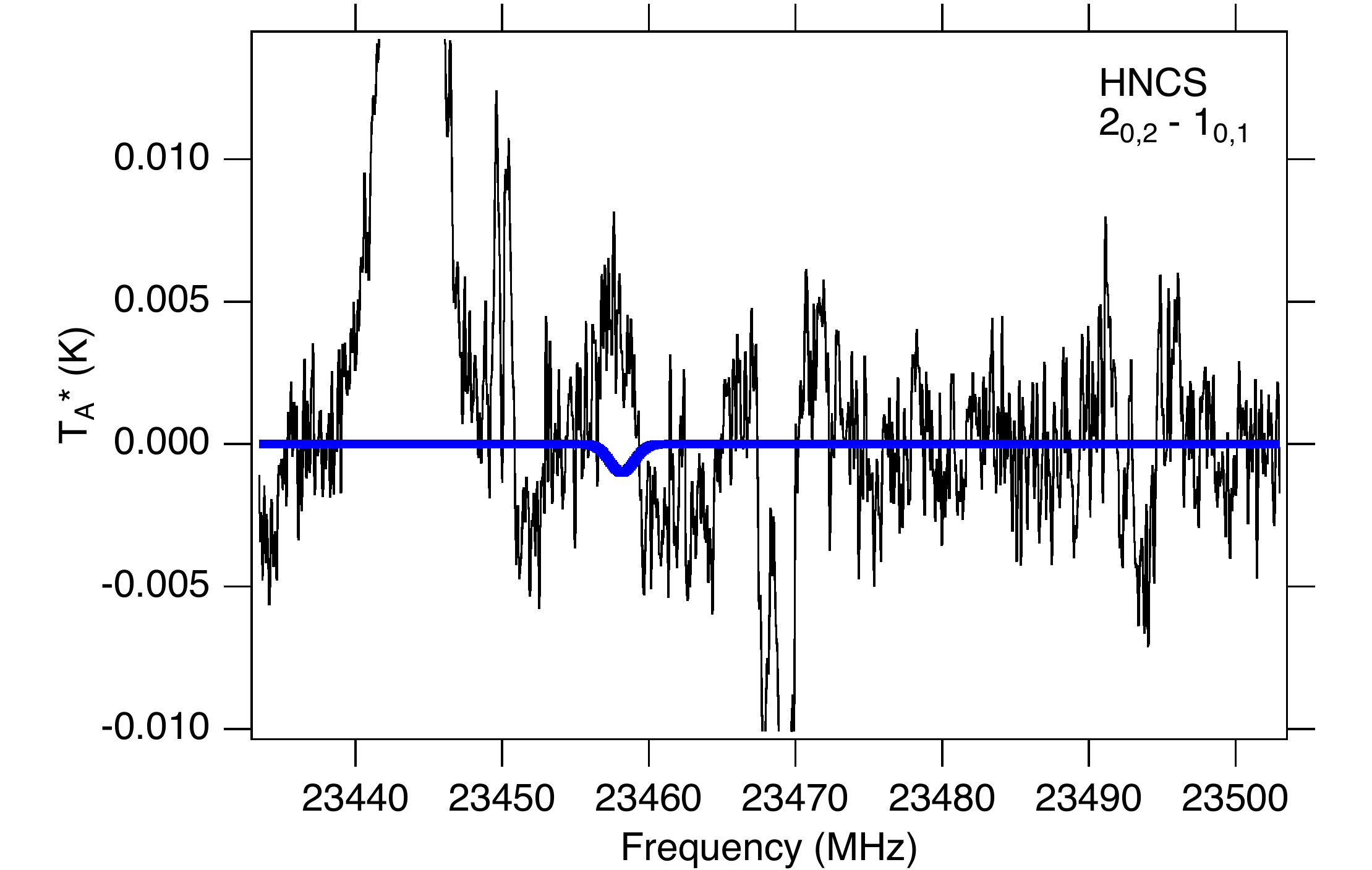}
\caption{Simulation of HNCS at previously-observed\cite{Halfen:2009it} column density and excitation temperature (blue trace, $\Delta V = 10$ km/s, $V_{LSR} = +64$ km/s) overlaid on PRIMOS observations at 23.5 GHz.}
\label{hncs}
\end{figure}

HNCS has been previously detected in Sgr B2(N) with a column density $N_T = 1.3(5) \times 10^{13}$ cm$^{-2}$ and a rotational excitation temperature $T_{ex} = 19(2)$\ K.\cite{Halfen:2009it}   Only the $J = 2_{0,2}- 1_{0,1}$ transition of HNCS at 23.5 GHz is accessible with PRIMOS since the next two $K_a=0$ lines, at 35187 and 46916 MHz, fall within small gaps in the survey coverage, while the $1_{0,1}- 0_{0,0}$ at 11729 MHz is contaminated with radio frequency interference (RFI).   While a simulation based on the previously-observed column density and excitation temperature predicts a weak absorption at the frequency of the 2--1 line, no absorption feature is visible (Figure \ref{hncs}).  Instead, a weak emission feature could potentially be assigned to HNCS.  While this assignment is only tentative, if correct it would indicate that HNCS is not described by a single excitation temperature, and in fact may be experiencing weak maser action in this low-$J$ transition. Such behavior has been reported previously in PRIMOS data toward Sgr B2(N): both carbodiimide (HNCNH) and methyl formate (\ce{CH3OCHO}) display weak masing.\cite{McGuire:2012jf,Faure:2014iu}

\subsection{HSCN}

\begin{figure*}
\centering
\includegraphics[width=\textwidth]{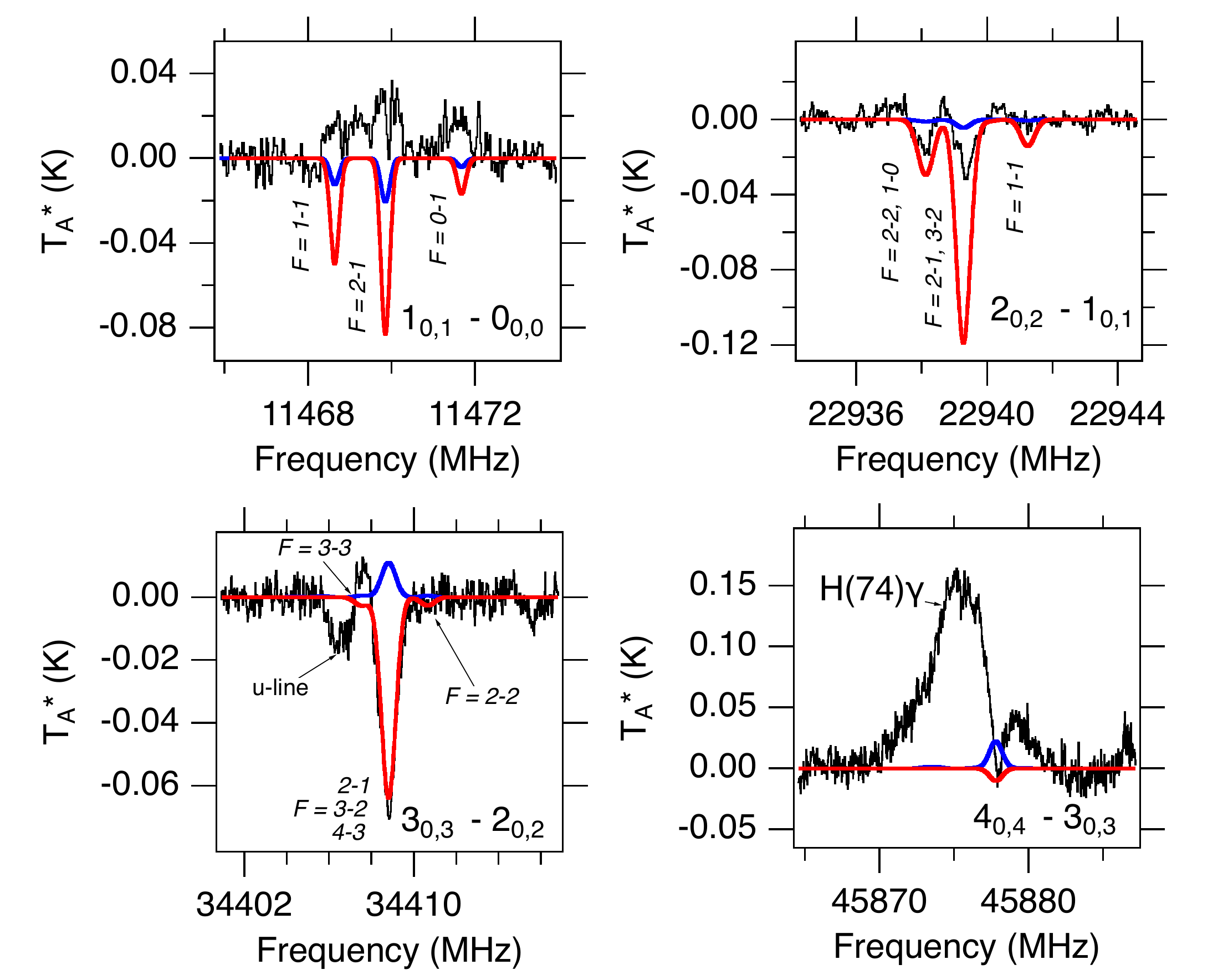}
\caption{Observed transitions of HSCN in the PRIMOS spectra in black, with simulations of HSCN based on the derived column density and excitation temperature from Halfen et al.\cite{Halfen:2009it} overlaid in blue, and the derived best-fit model from the PRIMOS observations ($T_{ex}=5$ K) overlaid in red.}
\label{hscn_obs}
\end{figure*}

HSCN has been previously detected in Sgr B2(N) with a column density $N_T = 3.3(2) \times 10^{13}$ cm$^{-2}$ and a rotational excitation temperature $T_{ex} = 18(3)$\ K.\cite{Halfen:2009it}  Unlike HNCS, four transitions of HSCN ($4_{0,4} - 3_{0,3}$, $3_{0,3} - 2_{0,2}$, $2_{0,2} - 1_{0,1}$, and $1_{0,1} - 0_{0,0}$) are covered by the available PRIMOS observations.  Using Eq. S1, and the previously-derived column density and temperature, these transitions are predicted to be detectable above the noise threshold of the PRIMOS observations.  As indicated in Fig. \ref{hscn_obs}, lines of HSCN are indeed detected at the expected frequencies, some with resolved hyperfine structure.  Because the hyperfine structure is known exactly from experiment, and is partially resolved in the astronomical data, our analysis explicitly treats this structure.  We find that a linewidth of 6.6 km/s, a column density of $N_T = 1.0 \times 10^{13}$ cm$^{-2}$, and a rotational excitation temperature of $T_{ex}$~=~5.0~K best reproduce the PRIMOS observations (Figure \ref{hscn_obs}; red).  This linewidth is significantly narrower than the previously-determined value of $\sim$25 km/s. The discrepancy is likely because \citet{Halfen:2009it} did not explicitly treat the hyperfine components of the transitions, as the components do not appear to be resolved in their higher-frequency observations.  Consequently, a much broader linewidth was derived.

The $4_{0,4} - 3_{0,3}$ transition of HSCN is seen in absorption superimposed on an emission feature from H(74)$\gamma$, while the remaining three transitions are unblended.  The  $3_{0,3} - 2_{0,2}$ is well-fit by our derived parameters, however the intensity of the $2_{0,2} - 1_{0,1}$ is over-predicted, and the $1_{0,1} - 0_{0,0}$ is observed in emission, rather than absorption.  Because the three higher-frequency transitions (at correspondingly lower continuum levels) are seen in absorption, the $1_{0,1} - 0_{0,0}$ line must be described by a separate excitation temperature greater than the continuum level at that frequency ($T_c>$125 K).  This behavior is strongly indicative of a weak maser, and indeed weak masing in the $2_{0,2} - 1_{0,1}$ transition would also account for the over prediction of the absorption depth in our single-excitation model.   We note that in the $1_{0,1} - 0_{0,0}$ emission line, narrow self-absorption features appear to be present.  If real, these features could be indicative of a spatially-distinct masing population embedded or behind colder, absorbing gas.  Owing to the limited data set, however, a detailed analysis is not feasible.

\subsection{HCNS and HSNC}

\begin{figure*}
\centering
\includegraphics[width=\textwidth]{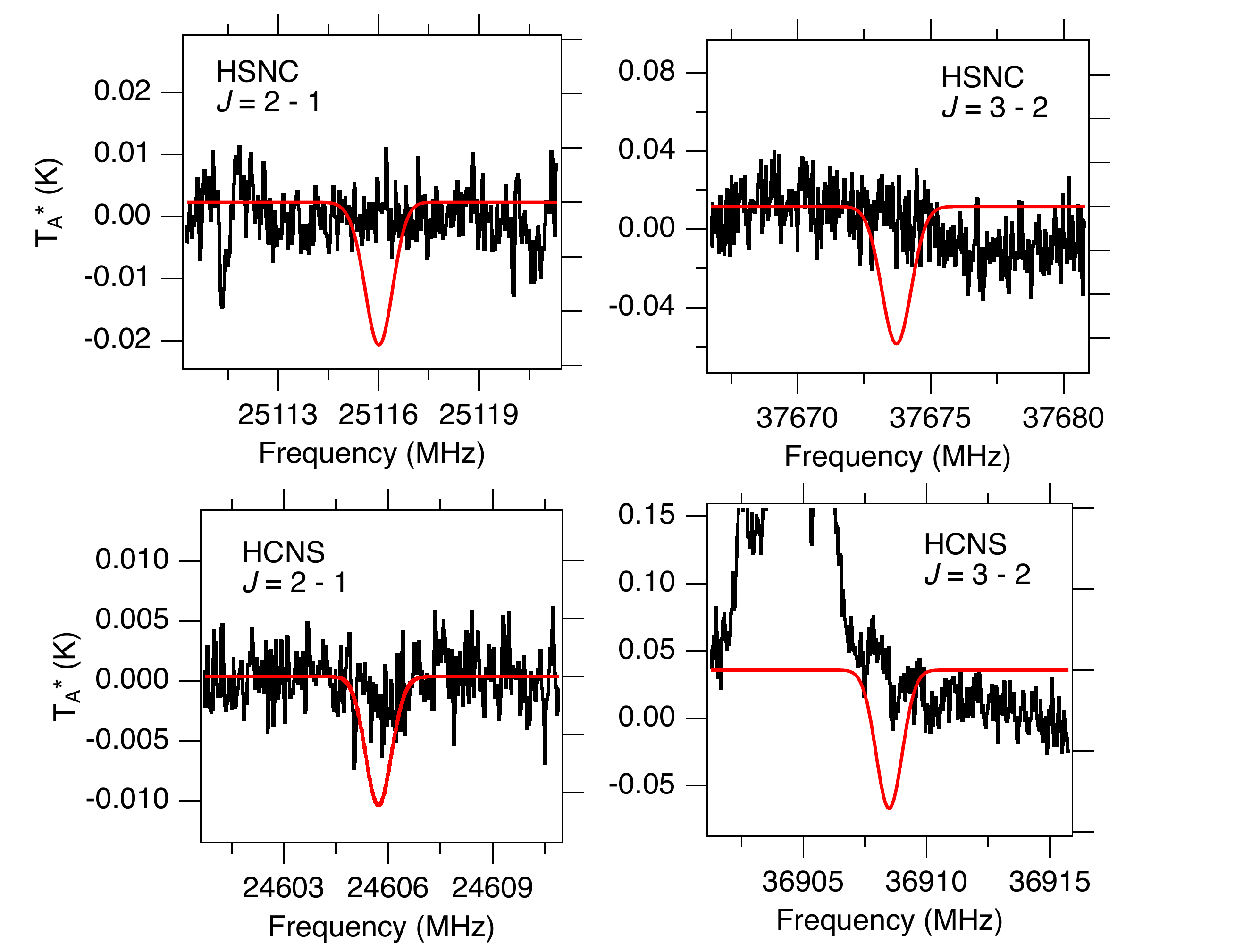}
\caption{Non-detections of HSNC and HCNS in PRIMOS spectra (black).  For illustration, red traces show simulations of the intensity profile of the transitions, in arbitrary units, for a linewidth of 10 km/s.}
\label{hcns-hsnc}
\end{figure*}

No compelling evidence is found for either HCNS or HSNC in the PRIMOS data (Figure \ref{hcns-hsnc}), yielding the upper limits given in Table \ref{cd_ulims} for a set of common conditions in the Sgr B2(N) environment.  In both cases, the $J = 1 - 0$ transitions of each fall within regions contaminated with RFI.  The PRIMOS observations are largely sensitive to cold, extended material, and thus the upper limits are correspondingly larger for warm, compact components.  If these isomers are present, they will likely only be detected if their ground-state transitions display masing enhancement like that of HSCN, or if high-sensitivity mm/sub-mm observations with instruments such as the Atacama Large Millimeter/sub-millimeter Array (ALMA) are undertaken.

\begin{table*}
\centering
\caption{Upper limits to column densities in PRIMOS observations toward Sgr B2(N) for four common conditions.}
\label{cd_ulims}
\begin{tabular}{c c c c }
\hline\hline
$T_{ex}$ (K)		&	Source Size ($^{\prime\prime}$)	&	\multicolumn{2}{c}{Upper Limit Column Density (cm$^{-2}$)}\\
			&								&	HCNS				&	HSNC			\\
\hline
8			&	20							&	$1.79 \times 10^{11}$	&	$1.12 \times 10^{12}$\\
18			&	20							&	$1.43 \times 10^{12}$	&	$2.07 \times 10^{13}$\\
80			&	5							&	$1.74 \times 10^{13}$	&	$2.13 \times 10^{14}$\\
140			&	5							&	$2.67 \times 10^{13}$	&	$3.02 \times 10^{14}$\\
\hline
\multicolumn{4}{l}{\emph{Note} -- We estimate the uncertainties in the upper limits to }\\
\multicolumn{4}{l}{be $\sim$30\%, largely arising from absolute flux calibration }\\
\multicolumn{4}{l}{ and pointing uncertainties.}
\end{tabular}
\end{table*}

\subsection{Astrophysical Implications}

The derived excitation temperature for HSCN agrees extremely well with other cold, extended molecules observed in PRIMOS, which are well described by $T_{ex} = 5 - 10$ K.\cite{Neill:2012fr,Hollis:2004oc,Loomis:2013fs}  We note that these excitation temperatures are dependent on the assumed continuum temperature, $T_c$, which in all these studies has been derived from the observations of Hollis et al.\cite{Hollis:2007ww}  The continuum structure in Sgr B2 is extremely complex, and observations with different beam sizes\footnote{The solid angle on the sky probed by the observations.}  (using other telescopes) and/or at higher-frequencies can be sensitive to different background continuum values.

The previously-reported column density and excitation temperature for HSCN systematically underpredict the observed PRIMOS features. In the cases of the 4$_{0,4} - 3_{0,3}$ and $3_{0,3} - 2_{0,2}$ transitions, they predict emission features, rather than the absorption features which are observed.  The difference is likely exclusively due to large variations in the continuum levels between the PRIMOS and ARO observations.  Nevertheless, a $T_{ex} = 19$ K is too high to completely account for the absorption seen in PRIMOS given the $T_c$ observed\cite{Hollis:2007ww}.  Conversely, the $T_{ex} = 5$ K derived here is likely too low to completely account for the previously-observed emission at higher frequencies; the lack of continuum information in the previous work prevents a rigorous comparison.  A full, quantitative radiative-transfer model\footnote{Such models explicitly treat the relative energy level population of molecules using both radiative and collisional excitation mechanisms.} would be required to properly reconcile the emission observed at mm-wavelengths with the absorption and potential masing seen in PRIMOS, but the necessary collisional cross-section information is not available.  Regardless, the PRIMOS observations are consistent with a cold, spatially-extended population of molecules distinct from species observed toward the hot core.

Such spatially-extended, low-excitation material in Sgr B2(N) is thought to be the result of bulk liberation of molecular ices from the surfaces of dust grains by the passage of low-velocity shocks.\cite{Hollis:2004uh,Chengalur:2003bf,RequenaTorres:2006ki} Indeed, several of the complex species detected toward this source are found exclusively to be cold and extended, and have not yet been detected in the compact hot core (e.g., $trans$-methyl formate, propanal, propenal,  propynal, and ethanimine).\cite{Neill:2012fr,Hollis:2004oc,Loomis:2013fs}  Previous work on HNCS and HSCN ratios in the ISM has suggested several gas-phase formation routes, proceeding through the cation precursor HNCSH$^{+}$.\cite{adande:561} Within pre-shock molecular ices, however, radical-radical reactions of simple species may be an important pathway for the evolution of more complex species.\cite{Garrod:2008tk,Garrod:2013id} HSCN and HSNC could also conceivably be formed in the condensed phase by the reaction of the SH radical with the CN radical, both known constituents of the ISM, before being liberated into the gas phase.\cite{McKellar:1940io,Jefferts:1970ld,Adams:1941tj,Neufeld:2012gz} Owing to its greater stability and larger barriers to isomerization, however, HSCN may be preferentially formed in space.  Radical-radical recombination reactions on grain surfaces are relatively energetic by interstellar standards, with the excess energy often stabilized by the grain surface acting as a third body.  Some of this initial energy, however, could easily drive population from HSNC to HSCN, where it is then trapped in the deeper potential well.

\section{Conclusions}

In the present investigation, the pure rotational spectra of HSCN, HCNS, and HSNC were recorded by a combination of chirped-pulse and cavity FTMW spectroscopy.  Rotational constants were obtained from fits to these spectra, and experimental $r_0$ structures were derived.  Using high-level quantum-chemical calculations, these structures were corrected for zero-point vibrational energy effects, and semi-experimental equilibrium $r_e^\text{SE}$ structures were determined for these three species, as well as for HNCS.  

Now that four isomers of the [H, N, C, S] system up to 37\,kcal/mol above ground have been characterized experimentally, even more energetic isomers may be within reach. On the singlet potential energy surface, Wierzejewska and Moc \cite{Wierzejewska:2003dd} calculate three ring molecules, $c$-C(H)NS, $c$-S(H)CN, and $c$-N(H)CS (i.e., where the hydrogen atom is bound to either of the heavy atoms forming a three-membered ring)  to energetically follow the four chains characterized here at roughly 45, 54 and 72\,kcal/mol above ground, respectively. Given that isomers of [H, N, C, O] as high as 84 kcal/mol above ground have been observed already,\cite{mladenovic:174308} any of the three [H, N, C, S]  ring isomers seem plausible candidates for future laboratory microwave searches. Even triplet species might be amenable to detection: the lowest triplet species, branched C(H)NS, is predicted at 63 kcal/mol followed by the bent chains HNCS and HCNS at 67 and 80\,kcal/mol, respectively.\cite{Wierzejewska:2003dd}

In the course of the present study, we have also searched the publicly-available PRIMOS centimeter wave survey of Sgr B2(N), and find no evidence for a cold population of HCNS or HSNC, and only a tentative detection of weak emission from HNCS.  Lines of HSCN are clearly observed, and evidence is found for weak maser activity in its $1_{0,1} - 0_{0,0}$ transition near 11469 MHz.  Future astronomical searches for HCNS and HSNC in molecule rich sources, however, are clearly warranted in the millimeter-wave regime. While the data obtained here are not sufficient to predict the millimeter wave spectra of these two species accurately enough for astronomical searches, they are indispensable in guiding  laboratory searches at still higher frequencies.

To further explore structural isomerism in analogous systems to the [H, N, C, O] family, comprehensive studies of molecules in which carbon and/or nitrogen are replaced with their heavier counterparts such as those of the [H, Si, N, O] and [H, C, P, O] families may be worthwhile.\cite{raghunath_JPCA_107_11497_2003,fu_CPL_361_62_2002} As a first step in this direction, the corresponding lowest-energy silicon and phosphorus analogs HNSiO and HPCO were recently detected by their pure rotational spectra.\cite{thorwirth_HNSiO_2015} Owing to the relatively small energy separation between isomers, and that there are very few experimental studies of these heavier analogs, members of these (seemingly simple) four-atomic molecular systems should provide a fertile testbed for further experimental study of molecular isomerism.

\section*{Acknowledgements}

This work was supported by NSF CHE 1213200 and NASA grant NNX13AE59G. B.A.M. is grateful to G.A. Blake for access to computing resources. S.T. acknowledges support from the Deutsche Forschungsgemeinschaft (DFG) through grants TH 1301/3-2 and SFB 956 and  the Regional Computing Center of the Universit\"at zu K\"oln (RRZK) for providing computing time on the DFG-funded High Performance Computing (HPC) system CHEOPS. The National Radio Astronomy Observatory is a facility of the National Science Foundation operated under cooperative agreement by Associated Universities, Inc.

%
%
%

\footnotesize{
\bibliography{HNCS_0710_2016} 
\bibliographystyle{rsc} 
}

\end{document}